\documentclass[12pt]{article}
\usepackage{latexsym}

\newcommand{\nc}{\newcommand}
\nc{\renc}{\renewcommand}

\nc{\etal}{\mbox{\it et al. }}
\nc{\ie}{{\it i.e. ~}}
\nc{\eg}{{\it e.g.}}

\renc{\thefootnote}{\arabic{footnote}}
\nc{\capt}[1]{{\bf Figure.} {\small\sl #1}}

\nc{\eqs}[2]{\mbox{Eqs.~(\ref{#1},\,\ref{#2})}}
\nc{\eq}[1]{\mbox{Eq.~(\ref{#1})}}

\nc{\figs}[2]{\mbox{Figs.~(\ref{#1},\,\ref{#2})}}
\nc{\fig}[1]{\mbox{Fig~.(\ref{#1})}}

\nc{\tag}[1]{\label{#1} \marginpar{{\footnotesize #1}}}
\nc{\mtag}[1]{\label{#1} \mbox{\marginpar{{\footnotesize #1}}}}
\renc{\baselinestretch}{1.2}
\newlength{\overeqskip}
\newlength{\undereqskip}
\nc{\be}[1]{\begin{equation} \mbox{$\label{#1}$}}
\nc{\bea}[1]{\begin{eqnarray} \mbox{$\label{#1}$}}
\nc{\Section}[2]{\section{#2}\label{#1}}
\nc{\Bibitem}[1]{\bibitem{#1}}
\nc{\Label}[1]{\label{#1}}

\nc{\eea}{\vspace{\undereqskip}\end{eqnarray}}
\nc{\ee}{\vspace{\undereqskip}\end{equation}}
\nc{\bdm}{\begin{displaymath}}
\nc{\edm}{\end{displaymath}}
\nc{\dpsty}{\displaystyle}
\nc{\bc}{\begin{center}}
\nc{\ec}{\end{center}}
\nc{\ba}{\begin{array}}
\nc{\ea}{\end{array}}
\nc{\bab}{\begin{abstract}}
\nc{\eab}{\end{abstract}}
\nc{\btab}{\begin{tabular}}
\nc{\etab}{\end{tabular}}
\nc{\bit}{\begin{itemize}}
\nc{\eit}{\end{itemize}}
\nc{\ben}{\begin{enumerate}}
\nc{\een}{\end{enumerate}}
\nc{\bfig}{\begin{figure}}
\nc{\efig}{\end{figure}}
\nc{\arreq}{&\!=\!&}
\nc{\arrmi}{&\!-\!&}
\nc{\arrpl}{&\!+\!&}
\nc{\arrap}{&\!\!\!\approx\!\!\!&}
\nc{\non}{\nonumber\\*}
\nc{\align}{\!\!\!\!\!\!\!\!&&}

\def\lsim{\; \raise0.3ex\hbox{$<$\kern-0.75em
      \raise-1.1ex\hbox{$\sim$}}\; }
\def\gsim{\; \raise0.3ex\hbox{$>$\kern-0.75em
      \raise-1.1ex\hbox{$\sim$}}\; }
\nc{\DOT}{\hspace{-0.08in}{\bf .}\hspace{0.1in}}
\nc{\Laada}{\hbox {$\sqcap$ \kern -1em $\sqcup$}}
\nc\loota{{\scriptstyle\sqcap\kern-0.55em\hbox{$\scriptstyle\sqcup$}}}
\nc\Loota{{\sqcap\kern-0.65em\hbox{$\sqcup$}}}
\nc\laada{\Loota}
\nc{\qed}{\hskip 3em \hbox{\BOX} \vskip 2ex}

\nc{\real}{{\rm I \! R}}
\nc{\Z}{{\sf Z \!\!\! Z}}
\nc{\complex}{{\rm C\!\!\! {\sf I}\,\,}}
\def\bigid{\leavevmode\hbox{\small1\kern-3.8pt\normalsize1}}
\def\id{\leavevmode\hbox{\small1\kern-3.3pt\normalsize1}}
\nc{\slask}{\!\!\!/}
\nc{\bis}{{\prime\prime}}
\nc{\pa}{\partial}
\nc{\na}{\nabla}
\nc{\ra}{\rangle}
\nc{\la}{\langle}
\nc{\goto}{\rightarrow}
\nc{\swap}{\leftrightarrow}

\nc{\EE}[1]{ \mbox{$\cdot10^{#1}$} }
\nc{\abs}[1]{\left|#1\right|}
\nc{\at}[2]{\left.#1\right|_{#2}}
\nc{\norm}[1]{\|#1\|}
\nc{\abscut}[2]{\Abs{#1}_{\scriptscriptstyle#2}}
\nc{\vek}[1]{{\rm\bf #1}}
\nc{\integral}[2]{\int\limits_{#1}^{#2}}
\nc{\inv}[1]{\frac{1}{#1}}
\nc{\dd}[2]{{{\partial #1}\over{\partial #2}}}
\nc{\ddd}[2]{{{{\partial}^2 #1}\over{\partial {#2}^2}}}
\nc{\dddd}[3]{{{{\partial}^2 #1}\over
	{\partial #2 \partial #3}}}
\nc{\dder}[2]{{{d #1}\over{d #2}}}
\nc{\ddder}[2]{{{d^2 #1}\over{d {#2}^2}}}
\nc{\dddder}[3]{{d^2 #1}\over
	{d #2 d #3}}
\nc{\dx}[1]{d\,^{#1}x}
\nc{\dy}[1]{d\,^{#1}y}
\nc{\dz}[1]{d\,^{#1}z}
\nc{\dl}[1]{\frac{d\,^{#1}l}{(2\pi)^{#1}}}
\nc{\dk}[1]{\frac{d\,^{#1}k}{(2\pi)^{#1}}}
\nc{\dq}[1]{\frac{d\,^{#1}q}{(2\pi)^{#1}}}

\nc{\cc}{\mbox{$c.c.$ }}
\nc{\cf}{cf.\ }
\nc{\erfc}{{\rm erfc}}
\nc{\Tr}{{\rm Tr\,}}
\nc{\tr}{{\rm tr\,}}
\nc{\pol}{{\rm pol}}
\nc{\sign}{{\rm sign}}
\nc{\bfT}{{\bf T }}

\nc{\cA}{{\cal A}}
\nc{\cB}{{\cal B}}
\nc{\cD}{{\cal D}}
\nc{\cE}{{\cal E}}
\nc{\cG}{{\cal G}}
\nc{\cH}{{\cal H}}
\nc{\cL}{{\cal L}}
\nc{\cO}{{\cal O}}
\nc{\cT}{{\cal T}}
\nc{\cN}{{\cal N}}
\nc{\rvac}[1]{|{\cal O}#1\rangle}
\nc{\lvac}[1]{\langle{\cal O}#1|}
\nc{\rvacb}[1]{|{\cal O}_\beta #1\rangle}
\nc{\lvacb}[1]{\langle{\cal O}_\beta #1 |}
\nc{\bb}{\bar{\beta}}
\nc{\bt}{\tilde{\beta}}
\nc{\ctH}{\tilde{\cal H}}
\nc{\chH}{\hat{\cal H}}
\nc{\lagr}{{\cal L}}
\nc{\dsub}[1]{\partial_{#1}}
\nc{\dsup}[1]{\partial^{#1}}
\nc{\av}[1]{\langle #1 \rangle}
\nc{\ordo}[1]{{\cal O}(#1)}
\nc{\conj}[1]{\overline{#1}}
\nc{\doo}[1]{\partial_{#1}}
\nc{\dop}[1]{\partial^{#1}}
\nc{\vev}[1]{\left\langle #1 \right\rangle}
\nc{\trivec}[3]{\left(\begin{array}{c} #1 \\ #2 \\ #3 \end{array}\right)}
\nc{\dmat}[4]{\left(\begin{array}{cc} #1 & #2 \\ #3 & #4 \end{array}\right)}
\nc{\1}{\aa}
\nc{\2}{\"{a}}
\nc{\3}{\"{o}}
\nc{\4}{\AA}
\nc{\5}{\"{A}}
\nc{\6}{\"{O}}
\nc{\al}{\alpha}
\nc{\g}{\gamma}
\nc{\Del}{\Delta}
\nc{\e}{\epsilon}
\nc{\eps}{\epsilon}
\nc{\lam}{\lambda}
\nc{\om}{\omega}
\nc{\Om}{\Omega}
\nc{\ve}{\varepsilon}
\nc{\mn}{{\mu\nu}}
\nc{\kp}{\kappa}
\nc{\vp}{\varphi}

\nc{\advp}[3]{{\it  Adv.\ in\ Phys.\ }{{\bf #1} {(#2)} {#3}}}
\nc{\annp}[3]{{\it  Ann.\ Phys.\ (N.Y.)\ }{{\bf #1} {(#2)} {#3}}}
\nc{\apl}[3]{{\it  Appl. Phys. Lett. }{{\bf #1} {(#2)} {#3}}}
\nc{\apj}[3]{{\it  Ap.\ J.\ }{{\bf #1} {(#2)} {#3}}}
\nc{\apjl}[3]{{\it  Ap.\ J.\ Lett.\ }{{\bf #1} {(#2)} {#3}}}
\nc{\app}[3]{{\it Astropart.\ Phys.\ }{{\bf #1} {(#2)} {#3}}}  
\nc{\cmp}[3]{{\it  Comm.\ Math.\ Phys.\ }{{ \bf #1} {(#2)} {#3}}}
\nc{\cqg}[3]{{\it  Class.\ Quant.\ Grav.\ }{{\bf #1} {(#2)} {#3}}}
\nc{\epl}[3]{{\it  Europhys.\ Lett.\ }{{\bf #1} {(#2)} {#3}}}
\nc{\ijmp}[3]{{\it Int.\ J.\ Mod.\ Phys.\ }{{\bf #1} {(#2)} {#3}}}
\nc{\ijtp}[3]{{\it Int.\ J.\ Theor.\ Phys.\ }{{\bf #1} {(#2)} {#3}}}
\nc{\jmp}[3]{{\it  J.\ Math.\ Phys.\ }{{ \bf #1} {(#2)} {#3}}}
\nc{\jpa}[3]{{\it  J.\ Phys.\ A\ }{{\bf #1} {(#2)} {#3}}}
\nc{\jpc}[3]{{\it  J.\ Phys.\ C\ }{{\bf #1} {(#2)} {#3}}}
\nc{\jap}[3]{{\it J.\ Appl.\ Phys.\ }{{\bf #1} {(#2)} {#3}}}
\nc{\jpsj}[3]{{\it J.\ Phys.\ Soc.\ Japan\ }{{\bf #1} {(#2)} {#3}}}
\nc{\lmp}[3]{{\it Lett.\ Math.\ Phys.\ }{{\bf #1} {(#2)} {#3}}}
\nc{\mpl}[3]{{\it  Mod.\ Phys.\ Lett.\ }{{\bf #1} {(#2)} {#3}}}
\nc{\ncim}[3]{{\it  Nuov.\ Cim.\ }{{\bf #1} {(#2)} {#3}}}
\nc{\np}[3]{{\it  Nucl.\ Phys.\ }{{\bf #1} {(#2)} {#3}}}
\nc{\pr}[3]{{\it Phys.\ Rev.\ }{{\bf #1} {(#2)} {#3}}}
\nc{\pra}[3]{{\it  Phys.\ Rev.\ A\ }{{\bf #1} {(#2)} {#3}}}
\nc{\prb}[3]{{\it  Phys.\ Rev.\ B\ }{{{\bf #1} {(#2)} {#3}}}}
\nc{\prc}[3]{{\it  Phys.\ Rev.\ C\ }{{\bf #1} {(#2)} {#3}}}
\nc{\prd}[3]{{\it  Phys.\ Rev.\ D\ }{{\bf #1} {(#2)} {#3}}}
\nc{\prl}[3]{{\it Phys\ Rev.\ Lett.\ }{{\bf #1} {(#2)} {#3}}}
\nc{\pl}[3]{{\it  Phys.\ Lett.\ }{{\bf #1} {(#2)} {#3}}}
\nc{\prep}[3]{{\it Phys\. Rep.\ }{{\bf #1} {(#2)} {#3}}}
\nc{\prsl}[3]{{\it Proc.\ R.\ Soc.\ London\ }{{\bf #1} {(#2)} {#3}}}
\nc{\ptp}[3]{{\it  Prog.\ Theor.\ Phys.\ }{{\bf #1} {(#2)} {#3}}}
\nc{\ptps}[3]{{\it  Prog\ Theor.\ Phys.\ suppl.\ }{{\bf #1} {(#2)} {#3}}}
\nc{\physa}[3]{{\it  Physica\ A\ }{{\bf #1} {(#2)} {#3}}}
\nc{\physb}[3]{{\it  Physica\ B\ }{{\bf #1} {(#2)} {#3}}}
\nc{\phys}[3]{{\it Physica\ }{{\bf #1} {(#2)} {#3}}}
\nc{\rmp}[3]{{\it  Rev.\ Mod.\ Phys.\ }{{\bf #1} {(#2)} {#3}}}
\nc{\rpp}[3]{{\it Rep.\ Prog.\ Phys.\ }{{\bf #1} {(#2)} {#3}}}
\nc{\sjnp}[3]{{\it Sov.\ J.\ Nucl.\ Phys.\ }{{\bf #1} {(#2)} {#3}}}
\nc{\spjetp}[3]{{\it Sov.\ Phys.\ JETP\ }{{\bf #1} {(#2)} {#3}}}
\nc{\yf}[3]{{\it Yad.\ Fiz.\ }{{\bf #1} {(#2)} {#3}}}
\nc{\zetp}[3]{{\it Zh.\ Eksp.\ Teor.\ Fiz.\  }{{\bf #1}  {(#2)} {#3}}}
\nc{\zp}[3]{{\it Z.\ Phys.\ }{{\bf #1} {(#2)} {#3}}}
\nc{\ibid}[3]{{\sl ibid.\ }{{\bf #1} {#2} {#3}}}
\nc{\rf}[1]{(\ref{#1})}
\nc{\nn}{\nonumber \\*}
\nc{\SM}{Standard~Model~}
\nc{\sm}{Standard Model}
\nc{\MP}{M_{\rm Pl}}
\nc{\tp}{t_{\rm Pl}}
\nc{\Appendix}
{
\setcounter{equation}{0}
\section*{Appendix}}
\nc{\gsm}{$SU(2)_L\times U(1)_Y$~}
\nc{\glr}{$SU(2)_L\times SU(2)_R\times U(1)_{B-L}$~}
\nc{\B}{\frac{M_Z^2}{M_{Z'}^2}}
\nc{\lrd}{\hbox{\raise1.3ex\hbox{$\leftrightarrow$}\kern-0.8em
\hbox{$\partial$}}}

\nc{\bra}{\langle}
\nc{\ket}{\rangle}
\nc{\st}[1]{\left| #1 \rangle\right.}
\nc{\ts}[1]{\left. \langle #1 \right|}
\nc{\tsh}[1]{{\tilde{ #1}}^\sharp}
\nc{\thc}[1]{{\tilde{ #1}}^\dagger}
\nc{\sh}[1]{{ #1}^\sharp}
\nc{\hc}[1]{{ #1}^\dagger}

\begin{document}
\begin{flushright} 
{\small TURKU-FL-P29-98}
\end{flushright}
\vspace{3cm}
\begin{center}
{\bf\Large Classical fields as coherent states and diffusion equations}
\end{center}
 
\begin{center}
Jukka Sirkka$^1$\\
and\\
Iiro Vilja$^2$\\ 
Department of Physics, University of Turku \\
 FIN-20014 Turku, Finland \\
\end{center}
\vspace{2cm}
\centerline{\bf Abstract}
\noindent 
We present a derivation of the effect of the classical
field configuration to the diffusion equations. Using the formalism of the
thermo field dynamics we propose a systematic and consistent way to treat the
classical background and to calculate it's effect at any order of the 
perturbation theory. We treat the classical field in genuine quantum field
theory formalism as a coherent state and show how the propagators and 
self energies are altered in this case.
\vfill
\footnoterule
{\small$^1$sirkka@newton.utu.fi,  $^2$vilja@newton.utu.fi}
\thispagestyle{empty}
\newpage
\setcounter{page}{1}

In statistical physics, the non-equilibrium phenomena are traditionally
described by the Boltzmann equations. Classically speaking, these
equations describe the transportation of particles in the background of
the statistical fluid, disturbed by the collisions between
particles. The Boltzmann equations can be derived in a variety of ways,
the most fundamental being from the Dyson-Schwinger (DS) equations of the
relativistic, statistical field theory.

If the non-equilibrium state is caused by spontaneous symmetry breaking
(SSB), a spatially and temporally non-homogeneous classical field
describes the propagation of the phase transition. In that case we would 
like to
have a way to calculate the effects of this SSB field to equilibration of
the particle gas. This can be done by working out the derivation from
the equations of the quantum field theory to the Boltzmann equations and
keeping in mind that the symmetry breaking field, i.e. the order parameter,
is a macroscopic classical field. From the Boltzmann equations one could
derive e.g. diffusion-like equations for the particle
densities. However, in the present paper, to make contact with the
studies of the Refs. \cite{de1,de4,de2,de3}, we by-pass the derivation of the
Boltzmann equations and write the diffusion equations directly from the
DS equations.

The proper way to deal with classical fields in quantum field theory is
to consider them as coherent states \cite{coherent}. This 
means that the field quanta propagate coherently, in constructive
interference, and form a macroscopic field. This state conserves its coherence
and the expectation value of the field evolves according to classical equations
of motion as far as the number of quanta in it remains large \cite{iv}.
In this proper treatment all phenomena, e.g. decay of the coherent 
state, are naturally incorporated to the theory. 

Recently there has been debate, how a classical field configuration, 
describing the propagation of the symmetry breaking phase transition, 
should be treated and what is its effect, especially in the context of the
particle gas out of equilibrium in the early universe \cite{de1,de2,de3}.
It seems not to be clear, how to incorporate the properties of the
domain wall into the quantum field theory. Various methods from more or less
heuristic \cite{de1} to a kind of semi-classical ones \cite{de4,de2} 
has been adopted leading to somewhat different results.

A well-argued technique to account the effects of the classical field
configuration is of especial importance when considering the
baryogenesis during the electroweak phase transition \cite{ewphase}.
Near the boundary of the broken and the symmetric 
phases the particle gas is in non-equilibrium state, thus fulfilling
one of the necessary Sakharov conditions \cite{saharov} for the
baryogenesis\footnote{The other conditions, baryon number violation and
CP-violation, can also be fulfilled: CP-violation through the usual
CKM-matrix based mechanism (or other mechanism in the case of the
extensions of the \SM) and baryon number violation due to the sphaleron
processes \cite{ewb:basics}.}.
The phase boundary acts then as a separator of quantum numbers. By various
methods the quantum numbers are converted to baryon number transported
to the broken phase and recognized as the baryon asymmetry of the universe.

We aim to show in the present letter how the classical field, which
describes the symmetry breaking phases in different regions in space,
can be adequately accounted using coherent states and thermo field
dynamics (TFD) \cite{ume}. The coherent state is the counterpart of classical 
field in the quantum field theory. On the other hand, TFD gives the way to
treat the non-equilibrium thermodynamics consistently \cite{tfd-and-ne}.

In the formalism of TFD, to account the doubling of the degrees of
freedom of the finite temperature field theory, one introduces the
operators $\xi_K$, $\tilde{\xi}_K$, $\sh{\xi}_K$ and $\tsh{\xi}_K$,
where $K\equiv (k_0, \vek k)$ labels the energy and the momentum. These
operators obey the commutation relations \be{comrel} [\xi_K, \sh{\xi}_P]
= (2\pi)^4 \delta^{(4)}(P-K), \,\,\,\, [\tilde{\xi}_K, \tsh{\xi}_P] =
(2\pi)^4 \delta^{(4)}(P-K) \ee and vanishing other commutators. These
are related to the operators $a_K$, $\tilde{a}_K$, $\hc{a}_K$ and
$\thc{a}_K$, which are used to construct the Heisenberg operator fields
and their tilde conjugates\footnote{ The finite temperature Hamiltonian
is unbounded from below\cite{landweert}. The ordinary fields represent
the particle-like excitations with positive energy while the tilde conjugated
fields represent hole-like excitations with negative energy.}, by the
Bogoliubov transformation
\bea{bog}
\left(\begin{array}{c} \xi_K \\ \tsh{\xi}_K \end{array} \right)  & = &
B \left(\begin{array}{c} a_K \\ \thc{a}_K \end{array} \right), \nonumber
\\
\left(\begin{array}{c} \sh{\xi}_K \\ -\tilde{\xi}_K \end{array}\right)^T  & = &
\left(\begin{array}{c} \hc{a}_K \\ -\tilde{a}_K \end{array} \right)^T B^{-1}.
\eea
Here the Bogoliubov matrix $B$ has a unit determinant and leaves the
form $\xi \sh{\xi} - \tsh{\xi}\tilde{\xi}$ invariant. Thus it is not 
uniquely defined. Intuitively, a simple choice is the so
called symmetric representation or $\alpha = 1/2$ -gauge, when
$\sh\xi_K$ is the hermitian conjugate of $\xi_K$ (which is not true in
general). This means that the Bogoliubov matrix satisfies the relation
$B^{-1} = \tau_3 B^T \tau_3$, where $\tau_3 = {\rm diag(1,-1)}$.  The
benefit of the symmetric representation is that the left- and right-handed
thermal vacua (see below) are conjugates. However, in the symmetric gauge there
appears to be some problems with the fermionic case \cite{henning}.

Another common choice for $B$ is the so called $\alpha = 1$ -gauge 
\cite{henning}, which we shall adopt in the present paper. Then
\be{alpha=1}
B= \left(\begin{array}{cc} 1+n & -n \\ -1 & 1\end{array}\right),
\ee
where $n$ is the bosonic number density. The benefit of this gauge is that 
also fermions can be treated consistently. Note further, that in finite
temperature the energy and momentum labels $k_0$ and $\vek k$ are not
related by a dispersion relation. This is a consequence of the broken
Lorentz-invariance in the thermal bath.

The Fock-space associated with the $\xi$-operators has the right-handed
vacuum state $\st{R}$, which is annihilated by the operators $\xi_K$ and
$\tilde{\xi}_K$; and the left-handed vacuum state $\ts{L}$, which is
annihilated by the operators $\sh{\xi}_K$ and $\tsh{\xi}_K$. The 
one-dimensional realization of the vacua, which can be readily adapted to the
more general case, is (in the $\alpha = 1$ -gauge)
\bea{vacua}
\st{R} & = & \exp\left(f\hc{a}\thc{a}\right)\st{0} \nonumber \\
\ts{L} & = & \ts{0}\exp\left(a\tilde a\right),
\eea
where $\st{0}$ is the vacuum state of $a$ and $\tilde a$ -operators and 
$f = e^{-\beta\omega}$ is the statistical weight. With these definitions,
statistical averages of observables correspond to the 
expectation values, 
\be{qwe1}
\Tr(\rho \,\hat O[A,\hc A]) = \ts{L}\hat O[a, \hc a] \st{R},
\ee
where $\rho = \exp(-\beta\omega \hat N)$ is the statistical operator 
and $\hat N = \hc A A$ is the number operator. 

At this point we make a generalization to the formalism of TFD by
shifting the operator $A$ with a parameter $\eta$ so that the new
statistical operator is
\be{statopfluk}
\rho_\eta = \exp\left (-\beta\omega(A^\dagger-\bar\eta)(A-\eta)\right ),
\ee
where bar denotes complex conjugate. The corresponding right-handed
vacuum of TFD is 
\be{tfdstatopfluk}
\st{R, \eta} = \exp\left(f\hc{a}\thc{a} +
(1-f)(\eta\hc{a}+\bar\eta\thc{a}-|\eta|^2)\right)\st{0}
\ee
and left-handed vacuum is left unchanged: $\ts{L,\eta} = \ts{L}$. Note that
the shift in \rf{statopfluk} generalizes to SSB field theory where one
shifts the symmetry breaking field with the vacuum expectation value. In the
zero temperature limit the state \rf{tfdstatopfluk} reduces to 
\be{coh}
\st{\eta} = \exp\left(\eta\hc{a}+\bar\eta\thc{a}-|\eta|^2\right)\st{0}.
\ee
Hence the expression \rf{tfdstatopfluk} can be regarded as a finite
temperature generalization of a coherent state \rf{coh}. This is in 
accordance with treating the symmetry breaking field as a coherent 
state. 

The $\xi$ operators, which annihilate
the vacua $\st{R, \eta}$ and $\ts{L, \eta}$ are now defined through 
\bea{bognew}
\left(\begin{array}{c} \xi \\ \tsh{\xi} \end{array} \right)  & = &
B \left(\begin{array}{c} a -\eta \\ \thc{a} -\eta \end{array} \right), 
\nonumber \\
\left(\begin{array}{c} \sh{\xi} \\ -\tilde{\xi} \end{array} \right)^T  & = &
\left(\begin{array}{c} \hc{a}-\bar\eta \\ -\tilde{a} + \bar\eta 
\end{array} \right)^T B^{-1}.
\eea

With these definitions, it is easy to calculate that (we are still 
considering the one-dimensional case)
\be{numberd}
\av{\hc{a}a} = 
\frac{\ts{L,\eta}\hc aa\st{R,\eta}}{\ts{L,\eta}R,\eta\ket } = n + |\eta|^2,
\ee
where
\be{nn}
n = \frac f{1-f},
\ee
i.e. number density is a sum of the contributions of the background
``field'' $\eta$ and quanta generated by the operator $A$. 

Changing to the field theory, one similarly postulates the left- and
right-handed vacua, which are annihilated by the raising ($\sh{\xi}_K$ and
$\tsh{\xi}_K$) and lowering ($\xi_K$ and $\tilde{\xi}_K$) operators,
respectively. These are related to the $a$-operators by the generalized
Bogoliubov transformation \rf{bognew} for every mode separately:
$a_K - \eta_K \goto \xi_K$. Here $\eta_K$ is the Fourier transform of the
background field $\phi(x)$, to be discussed below. 

The Heisenberg fields are constructed in two steps. First, one
introduces the so called generalized free field $\phi_K(x)$  
\cite{landsman} \footnote{Our notation for $\phi_K$ is
slightly misleading here, because it depends on energy and {\em
square} of the momentum $\vek k$.},
\be{gff}
\phi_K(x) = k^{-2}\int \frac{d^3p}{(2\pi)^3}\, 
\delta(|\vek{p}| - k)(e^{-ik_0 t + i\vek{p}\cdot \vek{x}}\,a_P + 
e^{ik_0 t - i\vek{p}\cdot \vek{x}}\,\hc a_P)|_{p_0 = k_0}
\ee
and corresponding tilde conjugated field $\tilde{\phi}_K$. These are
hermitian fields, but generalization to more complicated cases is
straightforward.  In the second step one decomposes the interacting
Heisenberg field $A(x)$ as \cite{landsman}
\be{I-field}
A(x) = \int_0^\infty \frac{dk_0}{2\sqrt{2 k_0}} \int_0^\infty
{d^3k\over (2\pi)^{3/2}} Z^{1/2}(K) A_K(x)
\ee
so that the field $A_K$ has a limit $\lim_{t\goto-\infty} A_K =
\phi_K$ in Lehmann-Symanzik-Zimmermann (LSZ) sense. Note that the
positive definite function $Z$ is a generalization of the wave function
normalization factor of the vacuum field theory. In principle it is fully
determined by the canonical commutation relations and the LSZ-condition
\cite{landsman}. However, in practise its calculation is a tremendous 
task.  

Perturbation theory is derived now from the finite temperature form of 
Gell-Man-Low -formula (GML) \cite{landsman}: as usual, the Feynman rules are 
derivable from it. In the presence of the
coherent state $\eta_K$ it generalizes for the two point function to 
\be{gml}
iD(x,x') \equiv \ts{L,\eta}T[ \vek{A}(x)\vek{A}^\dagger(x') ]\st{R,\eta}
= \ts{L,\eta} T[\Phi(x) \Phi^\dagger(x') U ] \st{R,\eta},
\ee
where 
\be{u}
U = T \exp\left(\int_{-\infty}^\infty\,dt\,\hat H_I(t)\right) 
\ee
with the interaction Hamiltonian $\hat H_I = \hat H[\Phi] - 
\hat H_0[\Phi]$.
Here the thermal Hamiltonian is expressed as
$\hat H[\Phi] = H - \tilde H$, where $H$ is the ordinary hamiltonian governing 
the time evolution of the Heisenberg fields and $\tilde H$ is its 
tilde conjugate. For tilde conjugation rules, see Ref. \cite{ume}.

Thermal doublet notation was introduced in Eq. \rf{gml}: 
\be{Phi}
\Phi (x)= \left ( \begin{array}{c} \phi(x) \\
                                   \thc{\phi}(x)
                   \end{array}\right ),
\ee
the field $\phi$ having the same form as the interacting field $A$, but
$A_K$ replaced by $\phi_K$. The interacting thermal doublet field
${\bf A}$ is defined  with a formula similar to \rf{Phi}.

The zeroth order propagator (which is $2\times 2$ -matrix) can be
calculated even without specifying the interaction Hamiltonian. By a
straightforward application of Eqs. \rf{bognew}, \rf{gff}, \rf{I-field}
and \rf{gml} one obtains
\be{d0}
D^{(0)}(x,y) = D(x, y)_{\eta = 0} - i \bar\phi_c(y)\phi_c(x) J \equiv
 D^{(0)}(x, x')_{\eta = 0} + \delta D(x, x'),
\ee
where $\phi_c$ is the classical field 
\be{phic}
\phi_c(x) = \int {d^4K\over (2\pi)^4} \theta(k_0) Z(K) \left(e^{-iKx}\eta_K +
e^{iKx}\bar\eta_K\right)
\ee
and $J$ is a constant matrix 
\be{J}
J = \left ( \begin{array}{cc} 1 & 1 \\
                              1 & 1
             \end{array}
    \right ).
\ee
The $D_{\eta=0}$ part of the propagator \rf{d0} has the same form as the usual
TFD zeroth order propagator without coherent state. However, it depends 
on $\eta$ through the functional dependence of $Z(K)$ on $\eta$. 

Note, that in order to use DS equations (to derive quantum transport 
equations) it is necessary to assume that the zeroth order propagator is 
on mass-shell, in the sense that 
\be{ms-cond}
(\Box_x + m^2) D^{(0)}(x, x') = \delta^{(4)}(x-x')\vek{1},
\ee
where $\vek 1$ is $2\times 2$ unit matrix. This implies that $\eta_K$
and $Z(K)$ are proportional to the mass shell delta function $\delta(K^2
- m^2)$. In what follows, we adopt this simple, although generally
illegitimate, approximation. This approximation is physically well
motivated if the particle mass is large enough compared to its (thermal) width,
i.e. $\Gamma \ll m$. This makes it reasonable to speak about
particles because the free path $1/\Gamma$ is long enough so that
they can be regarded as ''free'' between
interactions.  Note also, that because particle current can be
written in the form \footnote{Here we consider complex scalar field with
corresponding generalizations in the notations.}
\be{c1}
\bra J_\phi^\mu (x)\ket = \frac i2 \bra \tilde\phi (x)\lrd^\mu
\phi (x)\ket 
= - \frac 12 \left [ (\partial_x^\mu - \partial_y^\mu) 
D^{12}(x, y)\right ]_{x = y},
\ee
the particle number $N(x)$ in zeroth order reads 
\be{N}
N(x) \equiv \bra J^0_\phi(x)\ket = \int \frac{d^3k}{(2\pi)^3}
n(\omega_k) + \frac i2 \bar\phi_c(x)\lrd_0\phi_c(x),
\ee
where mass shell form of $Z(K)$ was used, so that $\omega_k = (k^2 +
m^2)^{1/2}$. Thus the Eq. \rf{N} produces the conventional result. 

Within the above framework, it is possible to derive the quantum transport
equations in the presence of the SSB field. Note that the dynamics of
the interacting fields, i.e. the Heisenberg equations, are not affected
by inclusion of the thermal coherent state. 
The thermal state (whether the coherent state included or not) 
contributes to the propagators only through the boundary 
conditions for the propagator.

Although the quantum transport equations (i.e. Boltzmann equations) can
be derived from the DS equations\footnote{For example, see
Ref. \cite{henning}.} using conventional procedures, we choose to step
directly to the diffusion equation, which is in many cases
sufficient. To begin with, the DS equation for the scalar
propagator matrix reads
\be{sd}
D(x, x') = D^{(0)}(x, x') + \int d^4y\int d^4z D(x, y)\Pi (y, z) 
D^{(0)}(y, x'),
\ee
where $\Pi$ is the self energy (which can be calculated
perturbatively) 
\be{self}
\Pi = \left ( \begin{array}{cc}
                  \Pi^{11} & \Pi^{12}\\ 
                  \Pi^{21} & \Pi^{22}
                 \end{array}\right ).
 \ee
There is also a similar equation to Eq. \rf{sd} with $D$ and $D^{(0)}$ 
interchanged in the right hand side. Note that for the simplest
case, constant background field $\phi_c$, the resumming of mass (in the
$\phi^4$ -theory) follows easily from Eq. (\ref{sd}).

Expressing the particle current as in Eq. \rf{c1} and using the DS
equations we obtain for the four divergence of the current
\bea{c3}
\dot n_\phi(x) + \nabla\cdot \vek j_\phi(x) &=& -\int d^4y \theta(x^0 - y^0)
\left [ \Pi^{21}(x, y) D^{12}(y, x) - \Pi^{12}(x, y) D^{21}(y, x)
\right. \nonumber\\
&-&\left. 
D^{21}(x, y) \Pi^{12}(y, x) + D^{12}(x, y) \Pi^{21}(y, x)\right ],
\eea
where $\bra J_\phi^\mu(x)\ket \equiv (n_\phi(x), \vek j_\phi(x))$. In 
deriving Eq. \rf{c3} the conditions
$D^{11} + D^{22} = D^{12} + D^{21}$ and 
$\Pi^{11} + \Pi^{22} = - \Pi^{12} - \Pi^{21}$ were used. 

Next we write corresponding equation for the fermion field $\psi$. 
The fermion propagator is defined through
\be{fp}
S(x,x') = -i \bra T[\Psi (x)\bar\Psi (x') ]\ket,
\ee
where $\Psi$ is the thermal doublet
\be{ff}
\Psi (x) = \left ( \begin{array}{c} 
                      \psi (x)\\
                      i{\tilde{\psi}^\dagger}(x)
                   \end{array}\right ),\ \ \ 
\bar\Psi = \left ( \bar \psi (x)\ \ \ \ - i\gamma^0\tilde{\psi}(x)
                   \right ).
\ee
The left- and right-handed vacuum states are generalized in an obvious way
to be the vacua with respect to the fermionic raising and lowering
$\xi$-operators.
Similarly as in the scalar case, using DS equations, one obtains
for the divergence of the particle current $\bra J_\psi^\mu \ket$,
\be{cf1}
\langle J_\psi^\mu (x)\rangle \equiv \langle \bar\Psi (x)^2\gamma^\mu 
\Psi (x)^1\rangle = \Tr [\gamma^\mu S(x, y)^{21}]_{x = y},
\ee
where trace is taken over Dirac indices,
\bea{cf3}
\dot n_\psi (x) + \nabla\cdot \vek j_\psi(x) &=& -
\Tr \int d^4y \theta(x^0 - y^0)
\left [ \Sigma^{21}(x, y) S^{12}(y, x) - 
\Sigma^{12}(x, y) S^{21}(y, x) \right. \nonumber \\
&-& \left. S^{21}(x, y) \Sigma^{12}(y, x) + 
S^{12}(x, y) \Sigma^{21}(y, x)\right ],
\eea
where $\Sigma$ is the self energy of the fermion field $\Psi$. 
Similarly as in the scalar case, we 
used the conditions $S^{11} + S^{22} = S^{12} + S^{21}$ and 
$\Sigma^{11} + \Sigma^{22} = -\Sigma^{12} - \Sigma^{21}$.

Next step would be to express the self energies in terms of the propagators
(using perturbation theory) and obtain self-consistent equations for the
particle currents. We do not perform this step in all its length, but 
choose to study only the effect of the inclusion of the coherent state to
the continuity equation. 

Because we do not consider fermion fields to have any coherent part, the
only direct contribution to Eq. (\ref{cf3}) appears through the scalar
propagator $\delta D(x, x')$. We write the self energies in the form
\be{sel}
\Pi = \Pi_{\eta = 0} + \delta\Pi\quad\mbox{and}\quad
\Sigma = \Sigma_{\eta = 0} + \delta\Sigma,
\ee
where all coherent state dependence, i.e. dependence on $\delta D$,
is included in $\delta\Pi$ and $\delta\Sigma$. These quantities can be
calculated perturbatively from the Feynman rules. The bosonic 
continuity equation reads 
\bea{o1}
\dot n_\phi (x) + \nabla \cdot \vek j_\phi (x) &+& \int d^4y
\left [ \Pi^{11}(x, y) D^{12}(y, x) - \Pi^{12}(x, y) D^{22}(y, x)
\right. \nonumber\\
&+&\left. 
D^{11}(x, y) \Pi^{12}(y, x) - D^{12}(x, y) \Pi^{22}(y, x)
\right ]_{\eta = 0}\nonumber\\
 &=& \gamma_{\phi,cl}(x),
\eea
where the coherent field contribution reads 
\bea{cl1}
\gamma_{\phi, cl}(x)  &=& -  \int d^4y
\left [ 
\delta \Pi^{11}(x, y) D^{12}(y, x)_{\eta = 0} +
\Pi^{11}_{\eta = 0}(x, y) \delta D^{12}(y, x)  \right. \nonumber\\
&-& \delta\Pi^{12}(x, y) D^{22}_{\eta = 0}(y, x) -
\Pi^{12}_{\eta = 0}(x, y) \delta D^{22}(y, x) \nonumber\\
&+& \delta D^{11}(x, y) \Pi^{12}_{\eta = 0}(y, x) +
D^{11}_{\eta = 0}(x, y) \delta \Pi^{12}(y, x) \nonumber\\
&-& \delta D^{12}(x, y) \Pi^{22}_{\eta = 0}(y, x) -
D^{12}_{\eta = 0}(x, y) \delta \Pi^{22}(y, x)
\left. \right ].
\eea
The corresponding  formula for the fermionic equation is
\bea{o2}
\dot n_\psi(x) + \nabla\cdot \vek j_\psi(x) &+& \Tr \int d^4y
\left [ \Sigma^{21}(x, y) S^{11}(y, x) - 
\Sigma^{22}(x, y) S^{21}(y, x) \right. \nonumber \\
&-& \left. S^{21}(x, y) \Sigma^{11}(y, x) + 
S^{22}(x, y) \Sigma^{21}(y, x)\right ]_{\eta = 0}\nonumber\\
&=& - \gamma_{\psi, cl}(x),
\eea
where
\bea{cl2}
\gamma_{\psi, cl}(x) &=& -  \Tr \int d^4y
\left [ 
\delta \Sigma^{21}(x, y) S^{11}(y, x)_{\eta = 0} +
\Sigma^{21}_{\eta = 0}(x, y) \delta S^{11}(y, x)  \right. \nonumber\\
&-& \delta\Sigma^{22}(x, y) S^{21}_{\eta = 0}(y, x) -
\Sigma^{22}_{\eta = 0}(x, y) \delta S^{21}(y, x) \nonumber\\
&-& \delta S^{21}(x, y) \Sigma^{11}_{\eta = 0}(y, x) -
S^{21}_{\eta = 0}(x, y) \delta \Sigma^{11}(y, x) \nonumber\\
&+& \delta S^{22}(x, y) \Sigma^{21}_{\eta = 0}(y, x) +
S^{22}_{\eta = 0}(x, y) \delta \Sigma^{21}(y, x)
\left. \right ].
\eea
The left hand sides of Eqs. (\ref{o1}) and (\ref{o2}) are the usual
terms of the diffusion equation after applying Fick's law 
$\vek j = - D \nabla n$. 
In the linear approximation, the last term in l.h.s produces the damping 
term $-\Gamma\,n$. 
The right hand sides of these equations implement the effect of the
background field $\phi_c$. 

One can apply Feynman rules to obtain perturbative expressions to the
integral terms in Eqs. (\ref{o1}) and (\ref{o2}).
For example, lowest non-vanishing perturbation calculation applied to \rf{cl2}
reproduces exactly the result of Ref. \cite{de2}.
In these papers CP- violating effects are considered at electroweak domain 
wall. There the classical field is inserted to the theory as an interaction
with fermion field through a Yukawa coupling $g_Y \bar\psi \phi\psi$ thus
producing CP-violating current at two insertion level. 
Indeed, the same result can be obtained by noting that at one-loop level 
with Yukawa-coupling the correction to the fermion self energy reads 
($\alpha,\ \beta = 1,\ 2$)
\be{apu}
\delta\Sigma(x, y)^{\alpha\beta} = 
-g_Y^2 S_{\eta =0}^0(x, y)^{\alpha\beta} \delta D(x, y)^{\alpha\beta},
\ee
and $\delta S = 0$. The right hand side of Eq. \rf{o1} reads in the one
loop approximation 
\bea{gg}
\gamma_\psi(x)  &=& - g_Y^2\int d^4 y \theta (x^0 - y^0) \left [ \phi_c (x)\phi_c^\dagger (y) -
\phi_c (y)\phi_c^\dagger (x)\right ] \nonumber\\
& &\times \Tr  \left [ S^0_{\eta = 0} (x, y)^{21}
S^0_{\eta = 0} (x, y)^{12} -
S^0_{\eta = 0} (x, y)^{12}S^0_{\eta = 0} (x, y)^{21}\right ].
\eea
This leads to the formula as given in Ref. \cite{de2}.

In the present letter we have proposed a method for consistent treatment of
classical fields in the framework of thermo field dynamics. In the method, a
classical field is accounted as giving a coherent contribution to the thermal
vacuum of the TFD. By using perturbation theory, the effects of the classical
field can be calculated systematically. In the lowest non-trivial order the
calculated contribution of the background field to the fermionic diffusion
equation reproduces the result of Ref. \cite{de2}. 
Because the inclusion of the coherent state to the thermal vacuum
does not affect the formalism of the
quantum field theory, all relevant aspects of the underlying model
are automatically incorporated. In particular, possible CP, baryon or 
lepton number violating effects of the model are naturally accounted. 
This is important e.g. in cosmological considerations, where classical 
field configuration, i.e. coherent state, represents the regions of broken 
symmetry. We will return to these aspects in a forthcoming paper.
\newpage

\end{document}